# Edge-dependent electronic and magnetic characteristics of freestanding $\beta_{12}$-Borophene Nanoribbons

Sahar Izadi Vishkayi and Meysam Bagheri Tagani[*]

Address: Department of Physics, Computational Nanophysics Laboratory (CNL), University of Guilan, Po Box:41335-1914, Rasht, Iran.

* Corresponding Author. Email: m_bagheri@guilan.ac.ir

# Edge-dependent electronic and magnetic characteristics of freestanding $\beta_{12}$-Borophene Nanoribbons

**Sahar Izadi Vishkayi and Meysam Bagheri Tagani**[*]

Address: Department of Physics, Computational Nanophysics Laboratory (CNL), University of Guilan, Po Box:41335-1914, Rasht, Iran.

Abstract- Nanoribbons produced from cutting $\beta_{12}$-Borophene sheet is investigated by density functional theory. The electronic and magnetic properties of Borophene nanoribbons are studied and found that all considered ribbons are metal which is in good agreement with recent experimental results. $\beta_{12}$-Borophene nanoribbons have a lot of diversity due to existence of 5 Boron atoms in a unit cell of the sheet. The magnetic properties of ribbons are strongly dependent on the cutting direction and the edge profile. It is interesting that a ribbon with a specific width can be a normal or ferromagnetic metal with magnetization in just one edge or two edges. The spin anisotropy is observed in some ribbons so that magnetic moment is not the same in both edges in antiferromagnetic configuration. The effect comes from the edge asymmetry of the ribbons and results in the breaking of spin degeneracy in the bandstructure. Our findings show that $\beta_{12}$ nanoribbons are potential candidates for next-generation spintronic devices.

## 1- Introduction:

Borophene, a single layer of Boron atoms, has been recently synthesized in ultra-high vacuum condition by two independent groups [1,2]. Initial reports were some different so that the borophene synthesized by Mannix et al. [1] was buckled whereas Feng and co-workers [2] presented two flat phases as $\beta_{12}$ and $\chi_3$ [3]. Later, it was demonstrated that the buckling observed in Ref.[1] can be attributed to the undulation of the first layer of substrate [4] and the synthesized borophene phase is $\beta_{12}$. Next analysis showed that the $\beta_{12}$ borophene is a metal [5] and the structure can host Dirac cone [6].

After two successful experimental syntheses of borophene sheets, a lot of theoretical research has been devoted to studying the borophene properties in two recent years [7-17]. Mechanical properties of several borophene sheets have been investigated and their ideal strength, ultimate strain and Young's modulus have been reported [15, 17-21]. Superconductivity [22-26] and thermal conductivity [12,27] of borophene sheets have been studied and predicted that their superconductivity can be modulated by the strain while their thermal conductivity is low. Furthermore, oxidized [10, 28] and hydride [29-31] borophene sheets have been examined and found that oxygen or hydrogen absorption reduces the anisotropy of the structure. It has been shown that the structural anisotropy of the borophene can lead to the direction-dependent current-voltage characteristics [32, 33].

Cutting two-dimensional (2D) structures along one-direction makes them as one-dimensional nanoribbons. Nanoribbons have electronic, magnetic, and optical properties which are different from 2D structures. It is well known that the zigzag-edge graphene nanoribbons are metal while the armchair-edge nanoribbons are metal or semiconductor with respect to the ribbons width [34, 35]. Garcia-Fuente et al. [8] studied borophene nanoribbons produced from *2Pmmn* and *8Pmmn* borophene sheets. They found that *8Pmmn* borophene nanoribbons are more stable and have more interesting properties. Nanoribbons can be non-magnetic or magnetic dependent on the cutting directions. In addition, the *8Pmmn* nanoribbon can be a metal or a semiconductor with respect to the cutting direction and its width. Meng and co-workers [36] investigated *2Pmmn* borophene nanoribbons and reported that the ribbons produced by cutting the sheet along x-direction are metal, whereas, the ribbons produced from y-direction can be magnetic. They also found that upon hydrogenation all nanoribbons become non-magnetic.

Zhong et al. [37] has reported successful synthesis of borophene nanoribbons on Ag(110) surface very recently. They observed several phase of the borophene ribbons like $\chi_3, \beta$ and $\beta_8$. Motivated by these experimental works on the borophene sheets and ribbons, we study the electronic and magnetic properties of $\beta_{12}$ borophene nanoribbons using density functional theory for the first time. Results show that all considered borophene nanoribbons are metal and the edge magnetization is dependent on the cutting direction. In addition, we observe that some ribbons are magnetic in just one edge. The spin anisotropy of the edge states is also observed that is attributed to the reconstruction of the edge. The electron density analysis reveals that the charge accumulation occurs in some edges which is consistent with recent experimental results [38].

The next section is devoted to simulation method. The simulation results are presented in section 3. We analyze binding energy, electron density, transmission channel, electron localized function, bandstructure, and magnetization of the considered ribbons in detail. Some sentences are given as a summary in the end of the article.

## 2- Simulation Details

All calculations were performed using density functional theory (DFT) implemented in SIESTA package [39]. The interaction between valance and core electrons was described by norm-conserved Troullier-Martins pseudopotentials [40] and Perdew-Burke-Ernzerhof (PBE) [41] generalized gradient approximation (GGA) was employed as exchange-correlation functional. Cut-off energy was *200Ry* and *100* k-points centered at Γ-point was used in direction where the ribbon is periodic. We considered *30 Å* vacuum to eliminate interlayer interactions. All ribbons were fully relaxed until force was converged to *0.001eV/Å*. 13 orbitals were employed for each Boron atom consisting of 2 sets of orbitals of *s* type, 2 sets of *p* type and 1 set of *d* type with cut-off radius of *2.8 Å*, *3.35 Å* and *3.35 Å*, respectively.

## 3- Results and Discussion

Figure (1) shows an optimized $β_{12}$-Borophene sheet with 5 Boron atoms in the unit cell. Lattice constants of the sheet are equal to *a=5.15 Å* and *b= 2.97 Å* which are in good agreement with previous works [2, 3, 5]. The sheet has uniform vacancies that compensate the electron deficiency of Boron atoms and stabilize the structure. Bandstructure of the freestanding borophene sheet indicates that the structure is metal like other reported borophene sheets [1, 2, 7, 13, 33]. Cutting the borophene along x- or y-direction gives rise to the formation of nanoribbons with different edge shapes. The edge profile and the width of ribbons are key factors in the electronic and magnetic properties of the ribbons, so one can expect that the cutting direction is very important. In this research, we focus on the two cutting directions: along x or y.

Borophene nanoribbons provided by the cutting of the sheet along x-direction is denoted by *NuvXBNR* which *N* is the number of Boron atoms in a row of the ribbon unit cell, and *u(v)* stands for the edge shape of a ribbon unit cell which can be composed of two Boron atoms (hereafter denoted by *A*) or three Boron atoms (hereafter denoted by *B*). It is obvious from Figure (1) that the ribbons with even *N* are *AB* and the ones with odd *N* are *AA* or *BB*. We consider ribbons with the width of $N = 10$ to $N = 15$ so that the maximum width of the ribbon is 21 Å. Analysis of the optimized *NuvXBNRs* shows that all structures are normal metal and no magnetization is seen in the edges. In addition, the ribbons are flat structures without buckling like $\beta_{12}$-Borophene sheet. Figure 2(a), 2(b) and 2(c) shows the structure of 13*AA*, 13*BB*, and 14*ABXBNR*, respectively.

For *NAAXBNRs*, the bonding lengths of the edge Boron atoms are 0.2 Å shorter than the bonding length in the sheet. Furthermore, the angle of the bonding between three Boron atoms next to the edge atoms is deviated from the line ($180^0$) and is equal to $167^0$ leading to the deformation of the hexagonal ring near the edge of the ribbon. The total energy per atom, $E = -\frac{E_{NuvXBNR}}{N}$, and the binding energy, $E = -\frac{E_{NuvXBNR} - NE_B}{N}$, of the structures are plotted in Figure 2(d) and 2(e), respectively, where $E_B$ stands for the energy of an isolated Boron atom. It is observed that *NAAXBNRs* have the lowest energy so that their binding energy is even lesser than (*N*−1)*ABXBNR*. The increment of the ribbon width gives rise to the increase of the binding energy. The electron density and the electron localized function (ELF) of the *NAAXBNRs* are shown in Figure 3. Figure 3(a) indicates the electron is accumulated in the edge of the ribbons which is in agreement with recent experimental results. Scanning tunneling microscopy results of Ref. [37, 38] showed that the electrons are accumulated in the boundary of the $\beta_{12}$ sheet. In addition, it has been recently reported that the edge of $\beta$-Borophene ribbons hosts more electrons than the body of the ribbon. The ELF also supports the above results so that the electrons are completely localized in the edge of the ribbons. Our findings show that the edge of Borophene ribbons is able to absorb atoms and molecules. Mannix at al. reported that their synthesized Borophene was partially hydrogenated [1] which can be attributed to the edge absorption with respect to our results.

*NBBXBNRs* are more stable than *NAAXBNRs*. The bonding length of the edge Boron atoms is 3 percent shorter than the sheet. In addition, bonding of the edge Boron atoms with Boron atoms next to the edge is stronger in the ribbon due to the shorter bond length. The electron density analysis, Figure (4) shows that the electrons are accumulated in the body of the ribbon dissimilar to *NAAXBNR*. By comparison with Figure 3(a), one can claim that the ribbons with the *A* kind edge are more inclined to absorb atoms and molecules via the edge. The bandstructure and the transmission channel per spin of *NAAXBNRs* and *NBBXBNRs* are plotted in Figure 5. It is well seen that all structures are metal. The transmission channel was computed using counting of the energy bands crossing a specific energy. The transmission channel can be considered as the transmission coefficient in low temperatures and perfect coupling. Figure 6 shows the electron density and the bandstructure of *NABXBNRs* which are composed of *A* and *B* types edges. As mentioned above, the electron accumulation is observed in the *A* type edge, similar to *NAAXBNR*. *AB* edge ribbons are also metal and their transmission channel increases by the increase of the ribbon width.

In the following, the ribbons obtained by cutting the Borophene sheet along y-direction are investigated. As it is shown in Figure 1, five Boron atoms of a $\beta_{12}$-Borophene unit cell have different x positions, numbered by 1 to 5 in Figure 1 so that there is a lot of variety for the ribbons. We name each ribbon as *NYuvBNR* where *u*, *v* = 1...5 and *u* and *v* shows the number of the Boron atom which is in bottom or top edge, respectively. *N* stands for the number of the Boron atoms in a unit cell of the ribbon and describes the width of the ribbon. We investigate the ribbons with $N = 20$ to $N = 25$ so that the maximum width of the ribbon studied here is 24 Å. We found that for each *N* there are three different edge configurations, therefore, 18 ribbons are studied in details. It is interesting to note that there are just two distinct configurations for ribbons created from 2*Pmmn* and 8*Pmmn* Borophene sheets [14] but here, we are faced with more diversity. The diversity leads to the more complexity of the $\beta_{12}$-Borophene nanoribbons along y-direction.

The total energy per atoms of *NYBNRs* is plotted in Figure 7. The total energy analysis shows that the *NYBNRs* can be magnetic in some widths, unlike *NXBNRs*. First, we study allotropes of 20*YBNR* and 25*YBNR* which have the same configurations. The increase of the ribbon width strengthens the stability of the ribbon which is clear with more energy of 25*YBNRs* than 20*YBNRs*. Although, some allotropes are magnetic, the ground state, 20*Y32BNR* and 25*Y32BNR*, is nearly nonmagnetic. The optimized structures of 20*YBNRs* are plotted in the supplementary information, Fig. S1. We find that the edge configuration, the bonding length, and the existence of fully occupied hexagonal lattices or hexagonal hole lattices are key factors in electronic and magnetic properties of ribbons. The electron localization function of *20YBNR* allotropes is plotted in Figure 8(a). It is clear that the edge significantly effects on the localization of electron. As it is seen in Figure 8(a), the electrons are localized in two edges of *20Y15* and *20Y12* leading to the magnetization of the structures. On the contrary, the electron density is distributed between two Boron atoms in the edge of *20Y32BNR*, especially, in the edge with hexagonal holes. We found that edges with hexagon holes are nonmagnetic, whereas, ones with fully occupied hexagonal rings are magnetic. A little electron localization is observed in the bottom edge of *20Y32* resulting in that energy of magnetic state of *20Y32* is a few *meV* more than nonmagnetic one. Highest electron localization is observed in *20Y21* and therefore, the energy difference between its magnetic and nonmagnetic states is more. We expect that if the ribbons growth on the substrate, the interaction between Boron atoms and the substrate becomes strong in some ribbons like *20Y21* and electron transfer between the ribbon and the substrate reduces the magnetization. For the magnetic state, we considered both ferromagnetic, two edges with same majority spin orientation, and anti-ferromagnetic, two edges with opposite majority spin orientation, configurations and found that they are degenerate. The spin density of allotropes of 20YBNR is depicted in Figure 8(b) in ferromagnetic configuration. Maximum of magnetic moment is equal to $0.58\mu_B$ for *20Y15* and to $0.7\mu_B$ for *20Y21*. A spin anisotropy is found in *20Y21* so that the magnetic moment is $0.7\mu_B$ in the top edge, whereas, it is $0.68\mu_B$ in the bottom edge. The anisotropy is directly dependent on the structural anisotropy and inequality of bond length in two edges of the ribbon. The spin anisotropy is amplified in other widths of the *NYBNR*. Bandstructure of *20YBNRs* is plotted in Figure S(4) which shows all ribbons are metal.

The total energy analysis shows that the ribbons with $N = 22, 23$, and 24 are magnetic independent of the edge profile. In the following, we analyze the most stable configurations of the mentioned ribbons and investigate the origin of the edge magnetization. The other considered ribbons are discussed in Supplementary Information. Figure 9(a) and 9(b) shows ELF and spin density in ferromagnetic configuration, respectively. ELF shows well that why these ribbons are magnetic. The electron localization is observed in both edges. The energy difference between magnetic and nonmagnetic states of 22*Y*51*BNR* is more than others because both edges are composed of fully occupied hexagonal lattices. It is clear that the electron localization is weaker in hexagonal hole lattices like *23Y13BNR*. The edge atoms are coupled to their neighbor's atoms anti-ferromagnetically in the y-direction and ferromagnetically in the x-direction. *22Y15* exhibits the strongest magnetization so the magnetic moment is $0.75\mu_B$ in both edges. The most spin anisotropy is observed in *24Y31* so that the magnetic moment difference between two edges is equal to $0.54\mu_B$, and upper edge has the higher magnetic moment. The spin anisotropy gives rise to spin splitting of the bandstructure in antiferromagnetic configuration, see Figure 10. Although one expects that the bandstructure becomes spin degenerate in anti-ferromagnetic configuration, the structural anisotropy of the edges breaks degeneracy and the bands become spin-dependent.

Results show that cutting $\beta_{12}$ sheet cannot induce a band gap in the structure, so all ribbons are metal which is in good agreement with recent experimental results [37]. Figure 10 describes bandstructure and density of states (DOS) in ferromagnetic and antiferromagnetic configurations. Each sharp peak of DOS is corresponding to an extremum of bandstructure or a flat band indicating strong electron localization. We found that the spin anisotropy significantly affects the bandstructure of antiferromagnetic configuration so that there are bands which are dependent on the spin orientation of edge atom with the highest magnetic moment. The bands are marked by a line in Figure 10. We assumed that the spin orientation of mentioned atom is spin-down in *23Y13*, and therefore, the bands are formed from spin-down electrons. About *24Y31*, the spin orientation is set to be spin-up for atom with the highest magnetic moment in antiferromagnetic configuration. Note that these bands are a direct consequence of spin anisotropy that were not reported in previous studies about Borophene nanoribbons [14].

We did not consider the role of substrate on the electronic and magnetic properties of nanoribbons. We expect that the substrate reduces the anisotropy of the structure and moderate the conductance of the structure. However, it was shown that $\beta_{12}$Borophene is metal and exhibits Dirac cone in the presence of the supported Ag [6]. On the other hand, recent experimental results show that the synthesized ribbons are metal and they are flat, therefore, we expect that the spin anisotropy reported in the article is a robust feature of some widths of the ribbon and it will be alive in the presence of the substrate. The spin anisotropy makes the $\beta_{12}$ nanoribbons a potential candidate for future spintronic and spin filtering devices. Unlike previous nanoribbons like Graphene nanoribbons, Silicene nanoribbons or Germanene nanoribbons, $\beta_{12}$-Borophene nanoribbons have a lot of diversity with individual characteristics which makes them an interesting candidate for next-generation electronic devices.

## 4- Summary

We have analyzed freestanding $\beta_{12}$-Borophene nanoribbons using density functional theory. The structural, electrical, and magnetical properties of ribbons are studied in details and found that the magnetization of ribbon is strongly dependent on its structural properties. Results show that all considered ribbons are metal and some of them can be magnetic. Magnetization is solely observed in ribbons prepared from cutting the Borophene sheet along y-direction *YBNR* and in some widths. Generally, *YBNRs* are more interesting so that some ribbons can be magnetic in one or two edges in specific widths. In addition, spin anisotropy is observed in *YBNRs*, so, one edge is more magnetic than the other. The spin anisotropy comes from asymmetry in edge of the ribbon. The spin anisotropy makes $\beta_{12}$-Borophene nanoribbons a potential candidate for spintronic applications.

**Figure Caption:**

Figure 1: The scheme of freestanding $\beta_{12}$-Borophene sheet and its unit cell is shown by the rectangular with *a* and *b* vectors (left panel). The band structure of the freestanding $\beta_{12}$-Borophene sheet (right panel).

Figure 2: The structure of a) *13AAXBNR*, b) *13BBXBNR* , and c) *14ABXBNR*. d)The total energy per atom and e) the binding energy of the considered structures.

Figure 3: a) The electron density and b) the electron localized function (ELF) of *11AAXBNR*, *13AAXBNR* and *15AAXBNR*.

Figure 4: The electron density of *11BBXBNR*, *13BBXBNR*, and *15BBXBNR*.

Figure 5: The band structure and the transmission channel of *11AAXBNR*, *15AAXBNR*, *11BBXBNR* and *15BBXBNR*.

Figure 6: The electron density (upper panels), the bandstructure and the transmission channel (lower panels) of *10ABXBNR*, *12ABXBNR* and *14ABXBNR*.

Figure 7: The total energy per atoms of *NYuvBNRs*.

Figure 8: a) The electron localization function and b) the spin density of *20YBNR* allotropes.

Figure 9: a) The electron localization function and b) the spin density of *22Y51*, *23Y13* and *24Y31*.

Figure 10: Spin-dependent bandstructure of a) *22Y51BNR* b) *23Y13BNR*, and c) *24Y31BNR* in anti-ferromagnetic (AFM) and ferromagnetic (FM) configuration. Density of states of each configuration is also drawn. Oblique lines show bands which are risen from the edge atom with highest magnetic moment. Dashed line is spin-down and solid line is spin-up.

Figure 1:

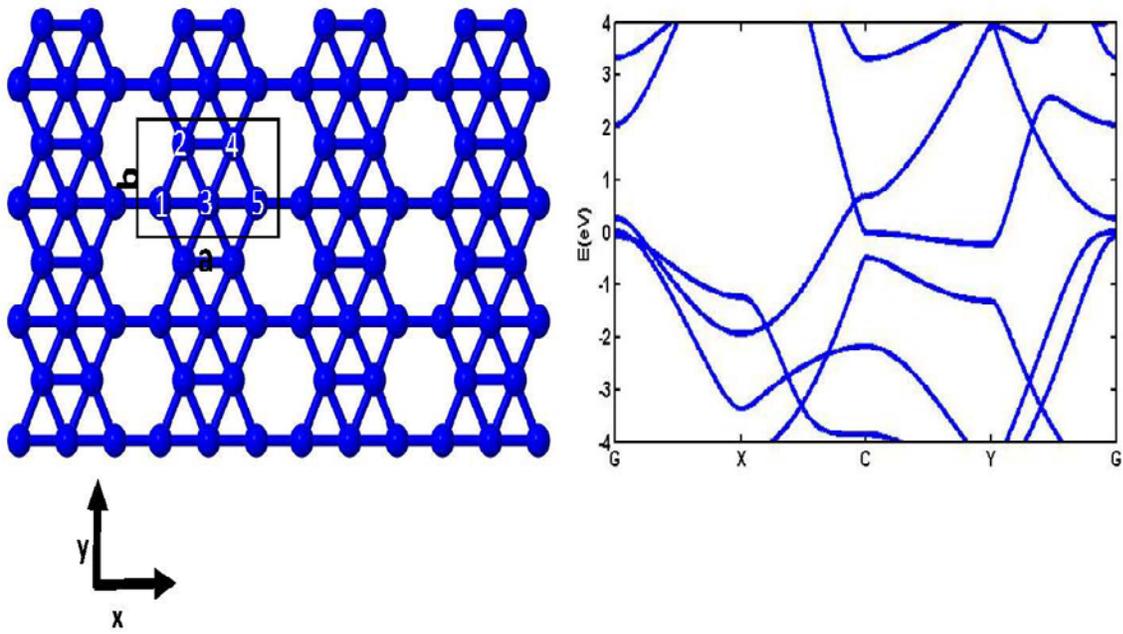

Figure 2:

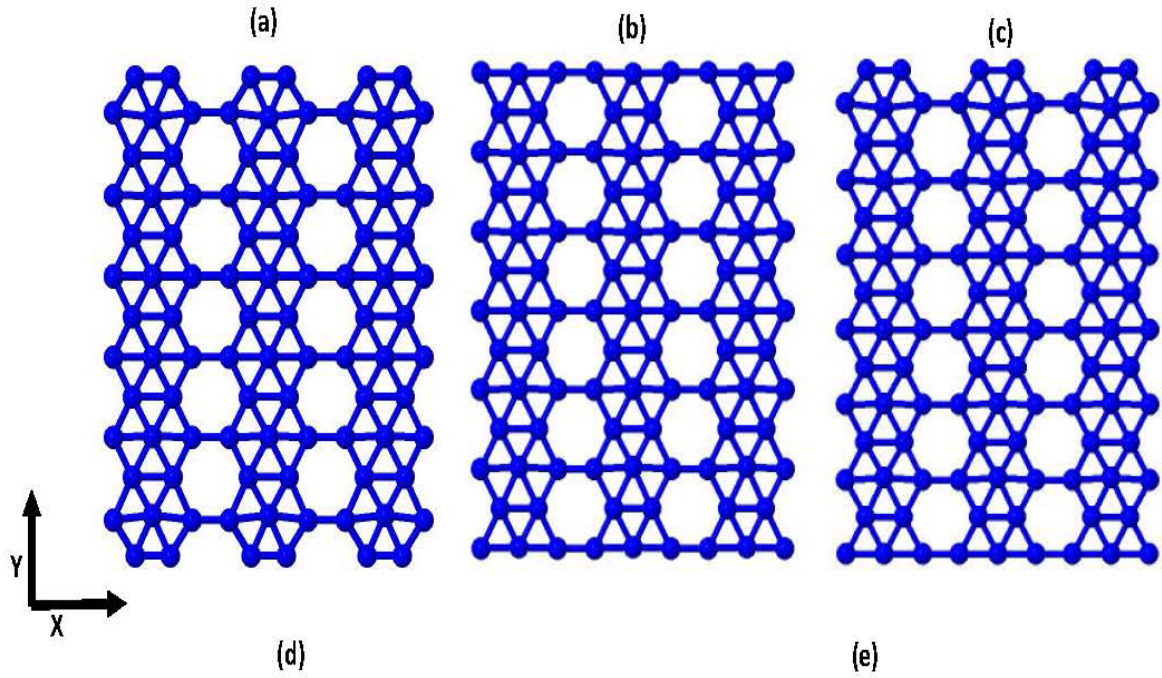

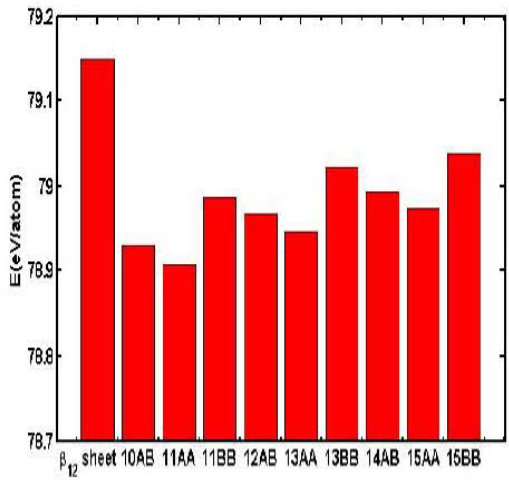
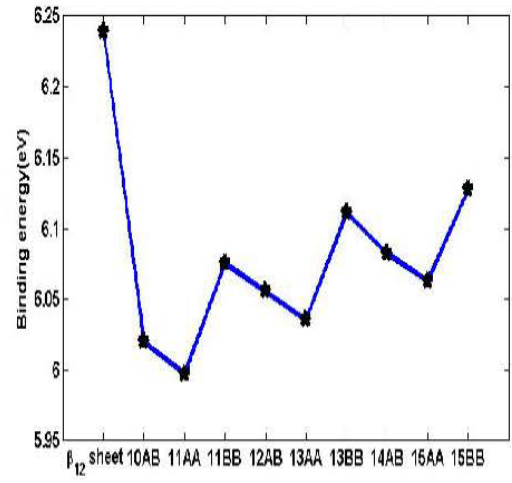

Figure 3:

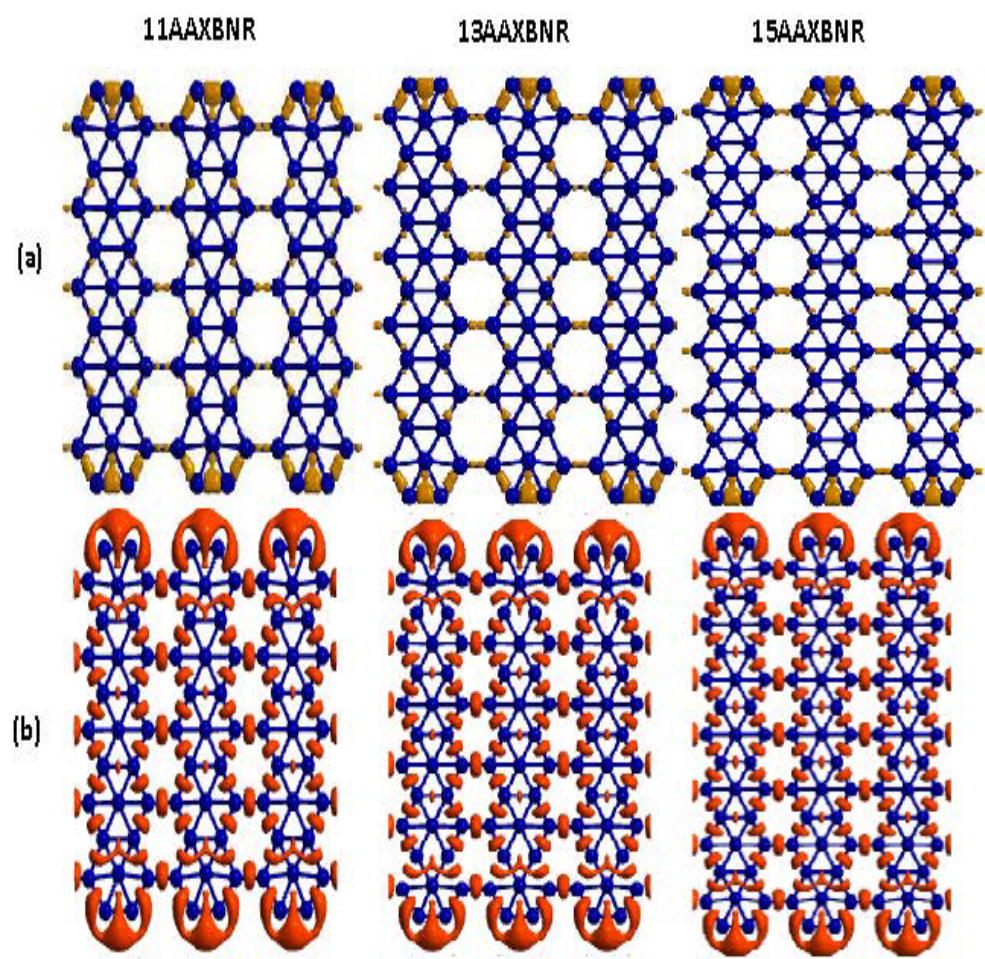

Figure 4:

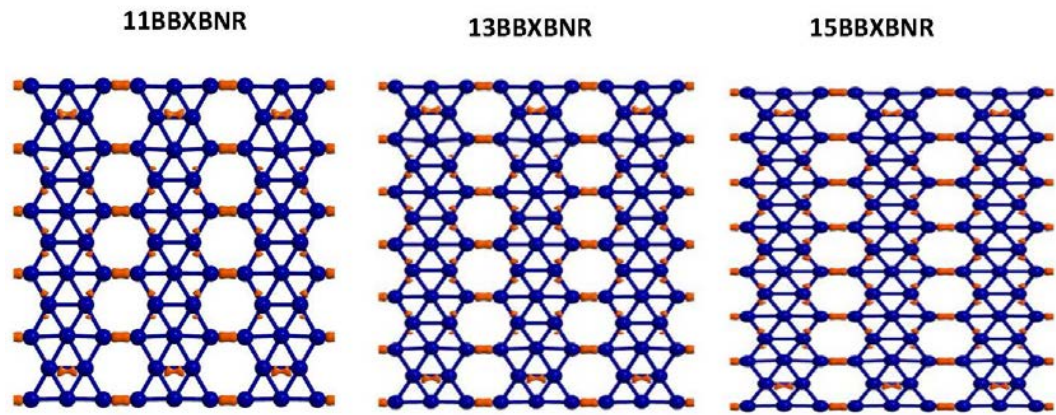

Figure 5:

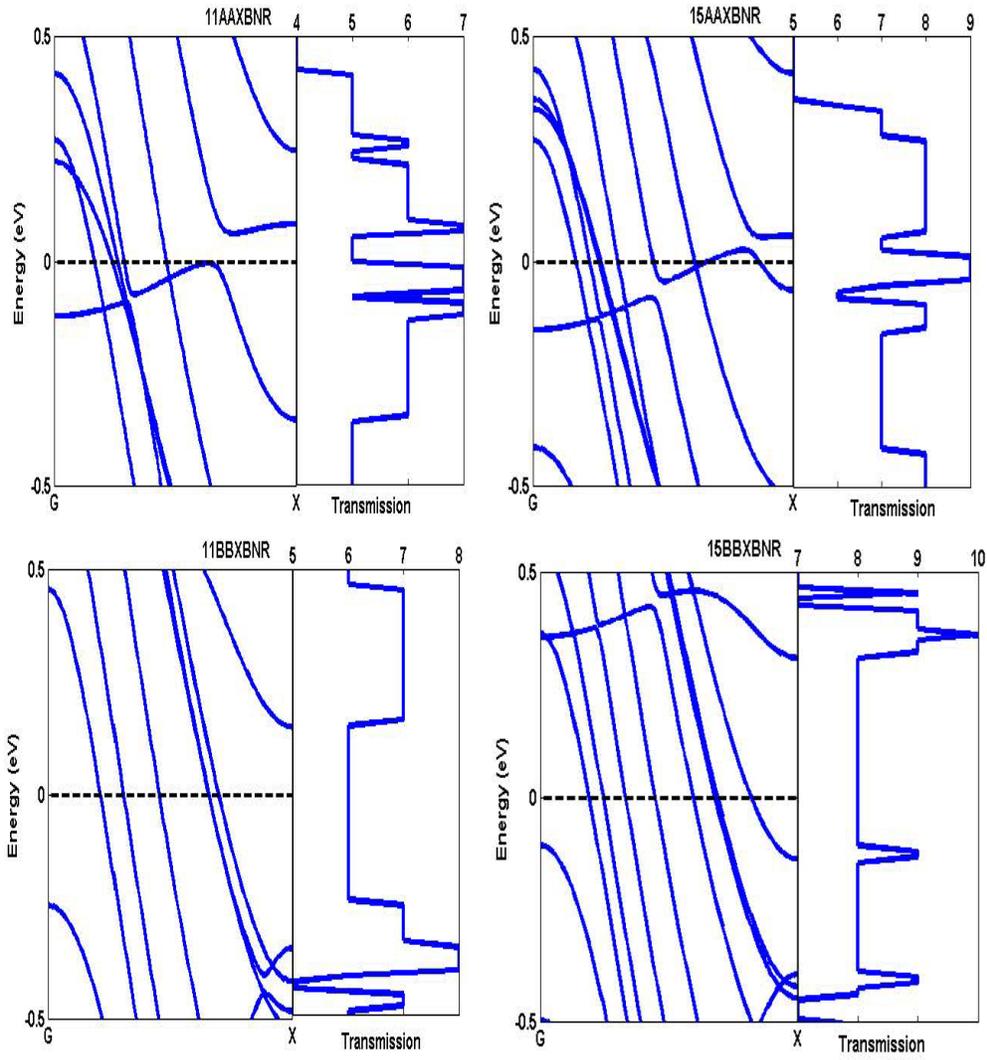

Figure 6:

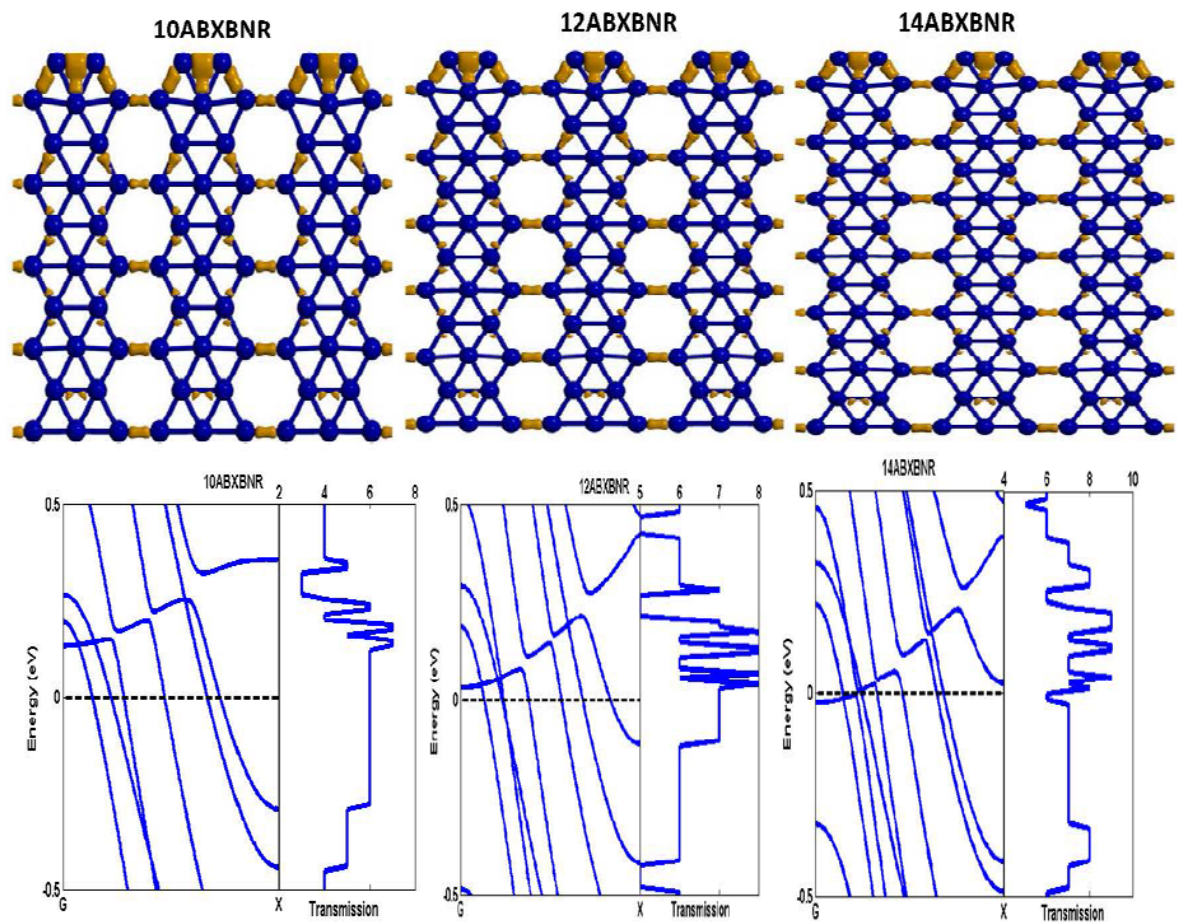

Figure 7:

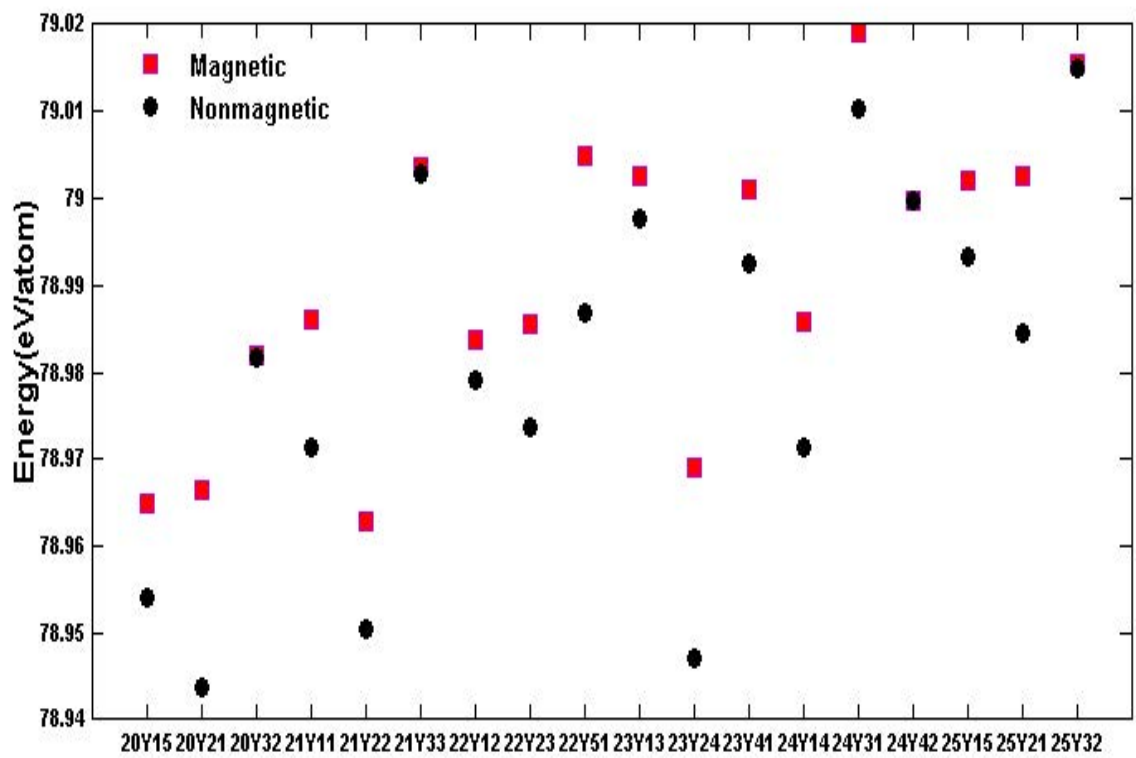

Figure 8:

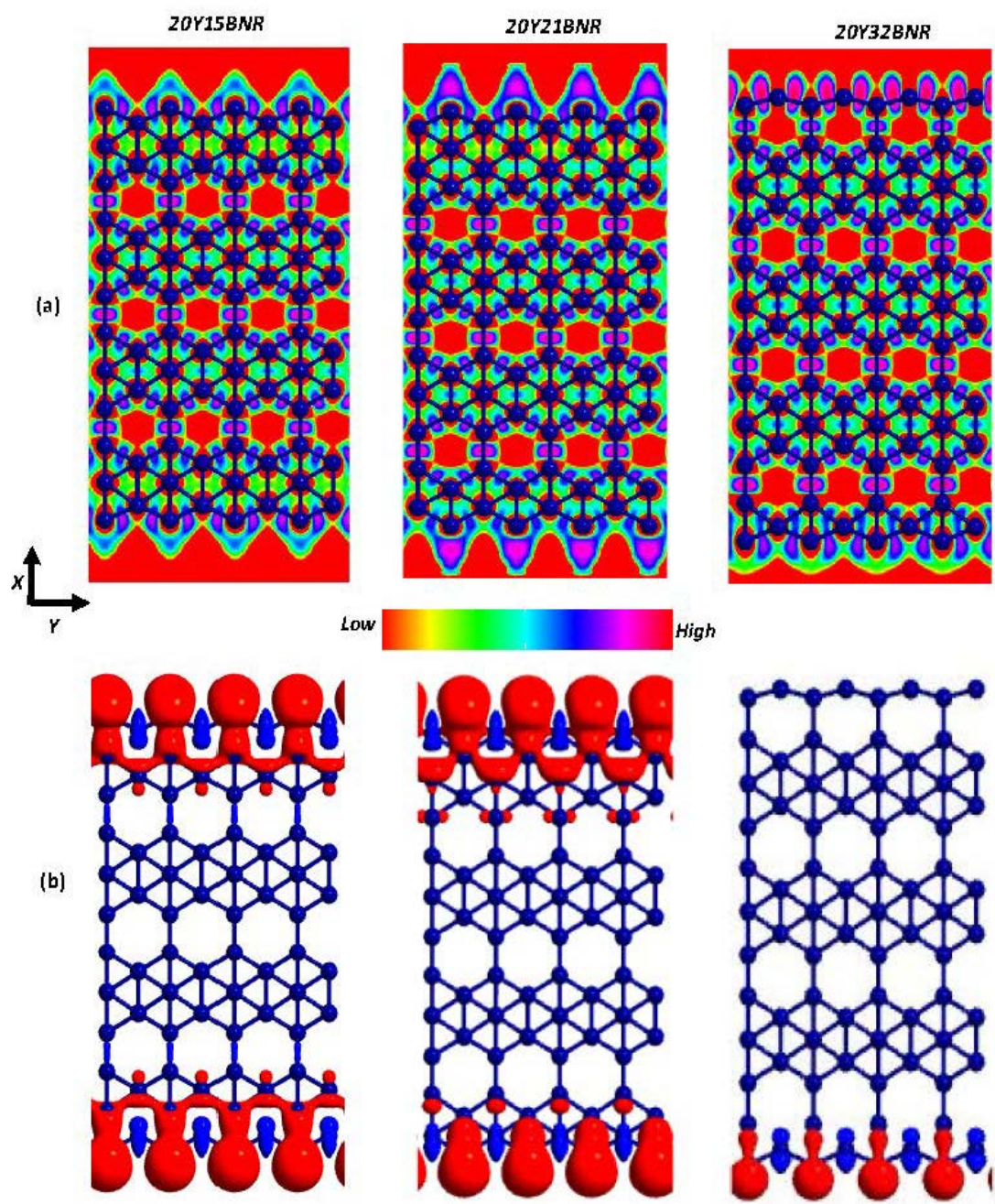

Figure 9:

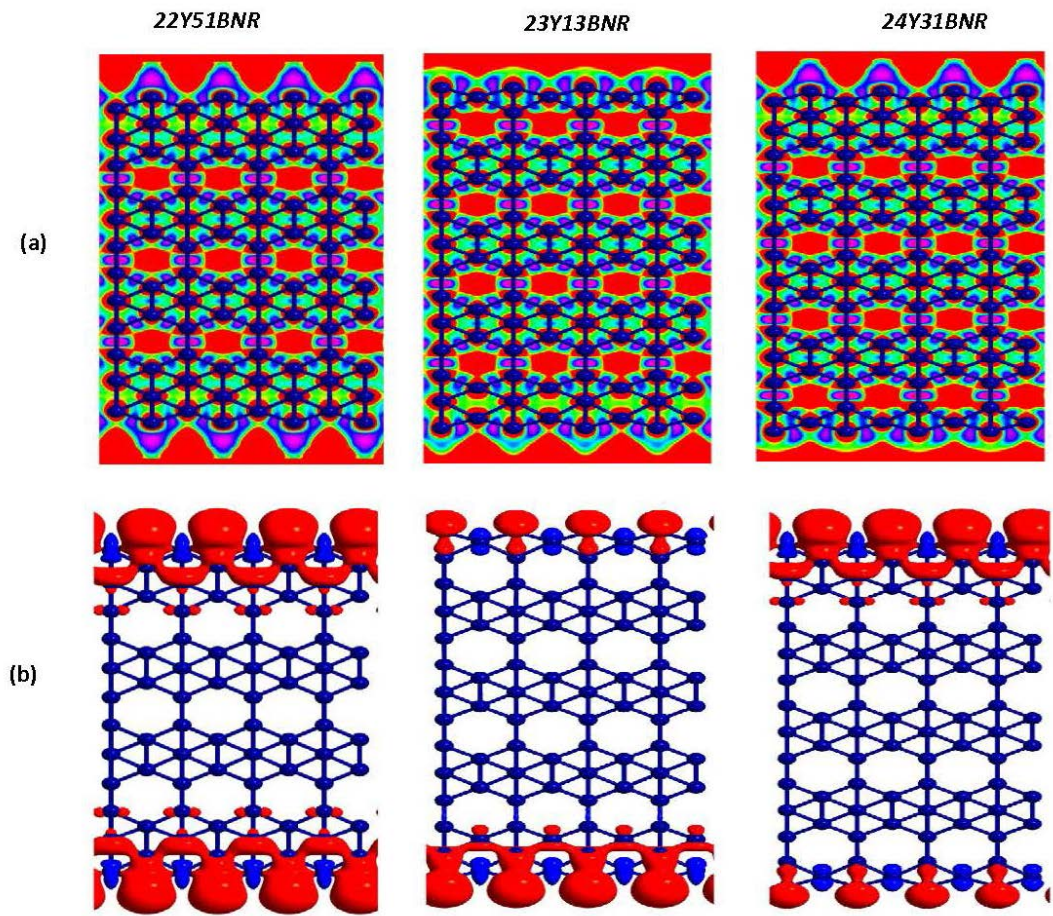

Figure 10 :

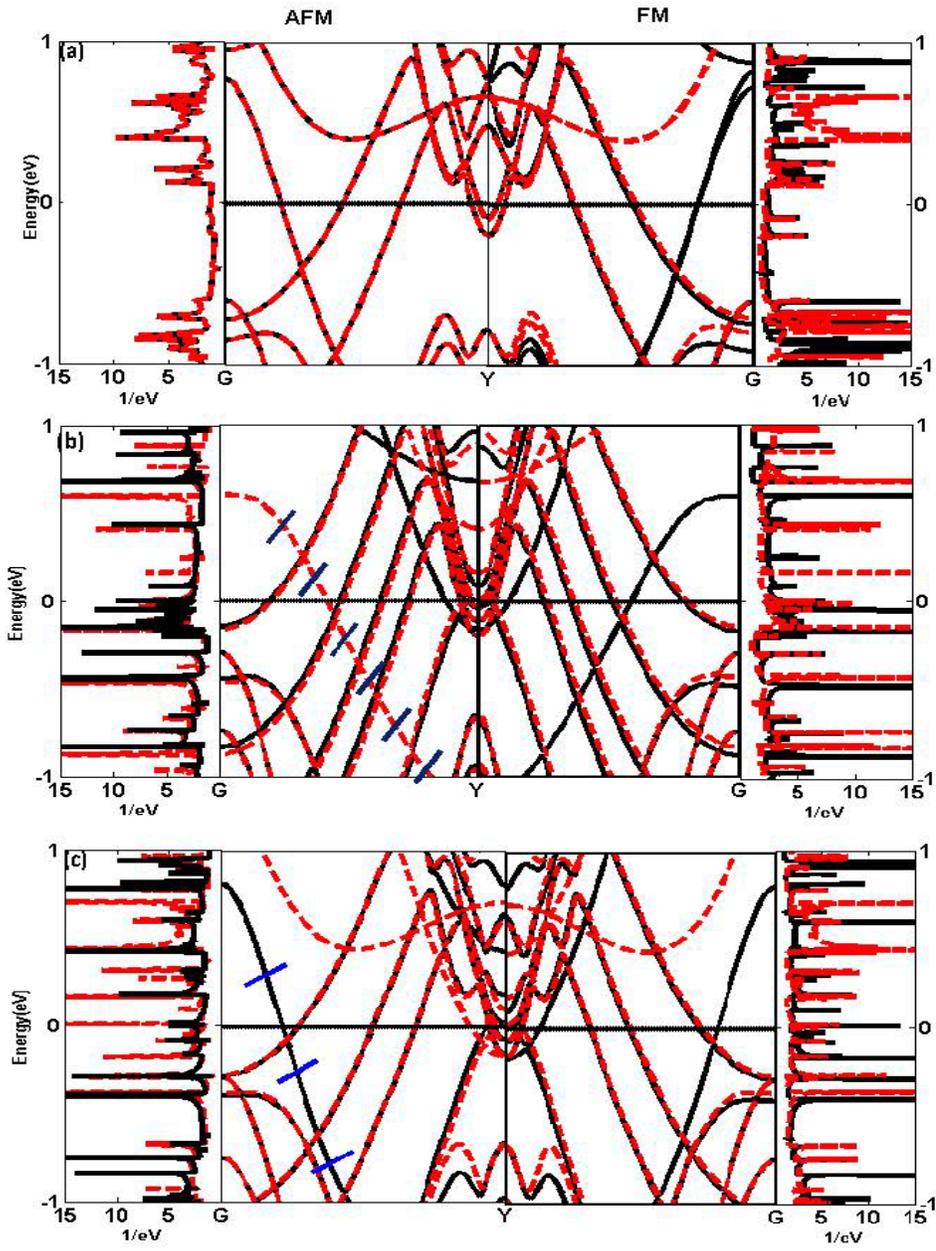